\documentstyle[preprint,tighten,aps,epsfig]{revtex}
\begin{document}
\begin{center}

\Large

{\bf Escape from a metastable well under a time-ramped force}

\large \vspace{0.5cm}
{\sl Julian Shillcock and Udo Seifert} \vspace{0.5cm}
\normalsize

 Max-Planck-Institut f\"ur Kolloid- und
Grenzfl\"achenforschung,\\ Kantstrasse 55, 14513 Teltow-Seehof, Germany
 \vspace{0.5cm}
 \end{center}
\begin {abstract}
Thermally activated
escape of an over-damped particle from a metastable well under the action of
a time-ramped force is studied. We express the
mean first passage time (MFPT)
as the solution to a partial differential equation, 
which we solve numerically for a model case. We discuss two approximations
of the MFPT,
one of which works remarkably well over a wide
range of loading rates, while the second is easy to calculate and can
provide a valuable first estimate. 
\end{abstract}
\def\beq{\begin{equation}}
\def\ee{\end{equation}}
\def\pcite{\protect\cite}
\def\pa{\partial}
\def\m{{\bf m}}
\def\q{{\bf q}}
\def\l{{\cal L}^\dagger}
\def\a{{\cal A}}

\def\lsim {\protect
\raisebox{-0.75ex}[-1.5ex]{$\;\stackrel{<}{\sim}\;$}}

\def\gsim {\protect
\raisebox{-0.75ex}[-1.5ex]{$\;\stackrel{>}{\sim}\;$}}

\def\lsimeq {\protect
\raisebox{-0.75ex}[-1.5ex]{$\;\stackrel{<}{\simeq}\;$}}

\def\gsimeq {\protect
\raisebox{-0.75ex}[-1.5ex]{$\;\stackrel{>}{\simeq}\;$}}

{\sl Introduction -- }
Thermally activated escape of
a particle from a metastable well, has found
numerous applications in a variety of systems\cite{z:haen90}.
Escape under the
additional action of a time-dependent force constitutes  a non-trivial
generalization. Several studies
\cite{z:fox89,z:jung89a,z:jung89,z:zhou90,z:casa94} have
been devoted to this problem for a {\it sinusoidal} force, which is particularly
interesting since this system shows stochastic resonance \cite{z:wies95a}.
The topic of the present paper is the effect of a {\it time-ramped} force on
the escape rate.
Apart from its fundamental significance,
 motivation to study this problem arises
from recent work on the dynamical strength of molecular bonds
\cite  {z:evan91,z:evan97d,z:izra97}.
 The
strength of a single bond can be measured experimentally using
atomic force microscopy where one in practice applies a
time-ramped force
\cite{z:evan91,z:evan97d,z:flor94,z:moy94,z:lee94}.
 Evans and co-workers \cite{z:evan91,z:evan97d}
 pointed out that  the
rupture strength of such a bond depends on the loading rate.
This behavior has been seen not only in
experiments but also in  simulations of a Langevin equation
\cite{z:evan97d,z:izra97} and
 molecular dynamics simulations
at very large loading rates \cite{z:izra97,z:grub96}.

For the general problem of diffusive escape from a metastable well under
a force that increases with time,
we express the  mean first passage time (MFPT)
  as the solution to a  partial
differential equation (pde). This exact approach differs from
 previous work that introduced an approximation based
on instantaneous decay rates \cite{z:evan97d,z:izra97}.
The numerical solution of the exact equation
is then
compared with both
this and another simple approximation. The first one works very well over
a large range of loading rates. The  second one can be calculated
easily and still gives a reasonable estimate which is never off more
than 30\% over the entire range of loading rates. 

{\sl MFPT as an exact solution to a 
pde -- } 
We consider the motion of an overdamped particle in
a one-dimensional
potential $V(x)$ under the action of a time-dependent
force $ f(t)$.
The Langevin equation for this
particle reads
\beq
\pa_t x = -V'(x) + f(t) + \zeta ,
\label{eq:la}
\ee
where the stochastic noise obeys the usual correlations
 $<\zeta (t)\zeta (t')> = 2 \delta (t-t')$. Throughout the paper
we measure energy in units of $k_BT$ and time and length such that
the diffusion coefficient becomes 1.
The potential $V(x)$ has a metastable well centered at $x=x_m$, a
saddle point at $x_s$ and an activation energy $Q\equiv V(x_s)-V(x_m)$.
 While our approach holds for any $f(t)$, we will
often specialize to a linear ramp
\beq
f(t)=\mu t
\ee
  with loading rate $\mu$.

 In the absence of the force ($f$=0), the MFPT
 $T(x)$ that a particle originally at $x$ needs to escape
from this well obeys the equation \cite{z:gard94}
\beq
\l T(x) \equiv (-V'\pa_x + \pa_x^2) T(x) = -1  .
\label{eq:t}
\ee
Here, $\l$ is the backward Fokker-Planck operator. The boundary conditions
are $T(x_l)=T(x_r)=0$, where for a metastable well
$x_l$ is far to the left of the minimum and $x_r$ far
to the right of the saddle point $x_s$. The explicit solution
$
T(x)$
as obtained by simple quadratures of
(\ref{eq:t}) is well known \cite{z:gard94,bc}.

In the presence of a time-dependent force,  relation
(\ref{eq:t})
 cannot be used without modification. To obtain
an exact equation for the MFPT, one can get rid
of
the time dependence of the right hand side in (\ref{eq:la}) by writing this
Langevin equation as a system of two equations
\begin{eqnarray}
\pa_t x &=& -V'(x) + f(\tau) + \zeta   , \nonumber \\
\pa_t \tau &=& 1 .
\label{eq:la2}
\end{eqnarray}
In this form, the problem formally corresponds to a
process  homogeneous in time
for the two variables $x(t)$ and $\tau(t)$. Therefore, one can apply the
standard equation for the MFPT which becomes
\beq
(\pa_\tau + f(\tau) \pa_x +\l) T(x,\tau) = -1  .
\label{eq:t2}
\ee
The boundary conditions in $x$ are as before $T(x_{l},\tau)=T(x_{r},\tau)=0$.
Since the deterministic variable $\tau$ does not diffuse, the equation
is first order in $\tau$. Therefore, one can specify only one boundary
 condition in $\tau$ which we choose as 
$T(x,\tau_e)=0$.  For these boundary conditions, the solution
$T(x,\tau)$ gives the MFPT that the process (4) starting for $t=0$ 
at $x$ and $\tau$ needs to reach either one
 of the following boundaries: (i) the particle reaches the absorbing
boundary at  $x_l$ or $x_r$;
or (ii) the
variable $\tau$ becomes $\tau_e$.
If 
$\tau_e$ is large enough, the number of events in which the
particle has not reached the absorbing boundaries at $x_l$ or $x_r$
but rather that at $\tau = \tau_e$ are
negligible compared to those in which it escapes over $x_l$ or $x_r$. 
Therefore, in the limit $\tau_e\to \infty$, the solution
$T(x,\tau = 0)$ gives the MFPT 
for the ramped problem. The
argument $\tau=0$ arises from the fact that the ramping starts 
with $f(t=0)=0$ at $t=0$.

{\sl Escape rate for constant force --}
Before we present numerical solutions of (\ref{eq:t2}), we discuss two
approximations for the special case of a linearly ramped force $f(t) =
\mu t$. To avoid a proliferation of symbols we use the same function
symbol $T$ for all MFPT quantities under a time-varying force, 
it being unambiguous from the arguments which case is being discussed. 
Both approximations use the solution to the
MFPT problem under constant force $f$, which we call $T_0(x,f)$, to construct 
approximate solutions for the MFPT under the time-varying force $f(t) =
\mu t$. The MFPT
$T_0(x,f)$ obeys
\beq
(f\pa_x + \l)T_0(x,f)=-1 ,
\label{eq:ts}
\ee which is easily solved by quadratures as \cite{z:gard94}
\beq
T_0(x,f)  = [\int_{x_l}^x {dy\over \Psi(y)}
\int_{x}^{x_r} {dy'\over \Psi(y')}
\int_{x_l}^{y'} dz \Psi(z) -
 \int_{x}^{x_r} {dy\over \Psi(y)}
\int_{x_l}^{x} {dy'\over \Psi(y')}
\int_{x_l}^{y'} dz \Psi(z)]/
\int_{x_l}^{x_r} {dy\over \Psi(y)} ,
\label{eq:t3}
\ee
with $\Psi(x)\equiv \exp(-V(x)+ fx)$. These integrals will be calculated
numerically.

A significant simplification occurs if one only requires the result
(\ref{eq:t3})  for large force-dependent activation barriers
\beq
Q(f)\equiv Q_0-f(x_s(f)-x_m(f)) >> 1.
\ee
A saddle point analysis of (\ref{eq:t3}) then  yields the well-known 
Kramers rate
\beq
T^K_0(f)=\tau_0(f)\exp [Q(f)] .
\label{eq:tk0}
\ee
The characteristic  time
\beq
\tau_0(f) \equiv {2\pi\over [V''(x_m(f))V''(x_s(f))]^{1/2}}
\label{eq:tau}
,
\ee
is the inverse
attempt frequency. $V''(x_{m,s}(f))$ is the curvature of the potential at the
minimum and the saddle point, respectively,
 whose locations depend on the force $f$.

We use the numerical calculation of the MFPT under constant force 
(\ref{eq:t3}), rather than the Kramers expression  (\ref{eq:tk0}),
to construct solutions for a 
time-varying force 
using two approximations. Both approximations thus obtained
are valid for all applied forces even when $Q(f)>>1$ no longer holds.

{\sl Self-consistent constant force (SCCF) approximation --}
The idea of this crude
approximation is that the escape is dominated by a typical
force
which is determied self-consistently.
In the solution (\ref{eq:t3}) for constant force we write
 $f= \mu T_0(x,f)$ which leads to the
self-consistent relation $T=T_0(x,\mu T)$. Solving this equation
  yields
the SCCF approximation $T_s(x,\mu)$. The  main virtue of this simple
approximation
is that it can be calculated  easily.

For large barriers, a further simplification can be achieved.
Using the Kramers expression (\ref {eq:tk0}), the SCCF
approximation
$T_s^K(\mu)$  follows from solving the implicit equation
\beq
T^K_s(\mu)={2\pi\over (V''(x_m)|V''(x_s))|^{1/2}}\exp (Q_0-\mu T^K_s (x_s-x_m))
\ee
where we notationally suppress the $f=\mu T^K_s$ dependence in $x_{m,s}=
x_{m,s}(f)$.

{\sl Adiabatic approximation --} A somewhat more refined approximation
has been introduced previously within the context of stochastic resonance
by Zhou et al \cite{z:zhou90} and for the ramped problem in Refs.
\cite{z:evan97d,z:izra97}. While the
SCCF approximation  focusses on the typical relevant force for the
escape, the adiabatic approximation  incorporates  history
dependence. It assumes  a time-dependent escape
rate $\nu (t)$. The probability that a given particle escapes
at time $t$ is then given by
\beq
p(t)=\nu (t) \exp [- \int_0^t dt'\nu(t')] . \label{eq:jcs1}
\ee

Equation (\ref{eq:jcs1}) says that the
probability of rupture is the product of two terms: the instantaneous
rupture rate $\nu(t)$ at a given time and the cumulative probability of survival
up to that time. This approximation 
rests on the assumption that the escape process takes no time. In this 
limit it becomes exact. For the linear  time-ramped process, one now replaces the 
time-dependence by a force-dependence via $f=\mu t$ which leads to the 
relation
\beq
p(f)=(1/\mu) \nu (f) \exp [-(1/\mu) \int_0^f df'\nu(f')] .\ee
Evans and Ritchie \cite {z:evan97d} discuss the conditions under which this
probability has a peak at non-zero $f^*$, which is defined as the
rupture force. Since within our approach, only the
MFPT is available, we will discuss this quantity
\beq
T_a(x,\mu) = (1/\mu) \int_0^\infty df
 \exp -[(1/\mu) \int_0^f df'\nu(f')] .
\label{eq:ta}
 \ee
Since $\nu(f)=1/T_0(x,f)$ with $T_0(x,f)$ from
(\ref{eq:t3}), the MFPT $T_a(x,\mu)$ in the adiabatic approximation
can be obtained by performing the two integrals numerically.

{\sl Numerical solution --}
In this section, we present numerical data which compare 
the MFPT obtained from  the two approximations with the numerical
solution of the exact equation (\ref{eq:t2}).

 Specifically, we  use
the  potential
\beq
V(x)\equiv (27/4) Q_0 x^2 (1-x)  .
\label{eq:V}
\ee
This potential exhibits a metastable minimum at $x_m=0$
with a saddle point at
$x_s=2/3$ and an activation energy $Q_0$.

Fig. 1 shows the MFPT  for two different values of the
activation energy
as a function of loading rate $\mu$
as calculated from
 (\ref{eq:t2}). For $\mu =0$, the MFPT approaches the zero-force value
 as obtained from (\ref{eq:t}). The inset shows that over
a large range $T\sim \mu^{-\alpha}$ where $\alpha \simeq 0.8$  is an effective
exponent masking an extended crossover from $\alpha =1$
(with logarithmic corrections)
to $\alpha =1/2$
as discussed below.

The quality of the two approximation is shown in Fig. 2, where we present the
ratio of the MFPT
\beq
R_{a,s}(x,\mu) \equiv T_{a,s}(x,\mu) / T(x,\mu)
\label{eq:R}
\ee
as calculated from either
approximation to the exact value.

The gross features of these data can be understood by identifying three
regimes \cite{z:evan97d,z:izra97}.
For small $\mu$, one expects naively that the MFPT deviates from
the value at zero force if $\mu \gsim \bar \mu$ where
\beq
\bar \mu \equiv Q(0)\exp[-Q(0)]/[\tau(0)(x_s-x_m)] .
\ee
While this (exponentially small!) loading rate also signifies the
range where the SCCF starts to deviate from the exact data, the adiabatic
approximation holds well even for larger loading rates.
This is remarkable since
as shown in the
Appendix
systematic perturbation theory  reveals that both approximations fail to
reproduce the
$O(\mu)$ correction to the zero force result for a general potential. It
seems that only in the additional limit of a large
activation energy, $Q(0)>>1$, is good agreement between the exact result and
the adiabatic approximation restored. A mathematical proof of this
feature is beyond the scope of this paper.

 For large $\mu$, both the potential and
the diffusion can
be ignored and then
the ratio displayed in Fig. 2 can be calculated analytically. Even
though the process in this regime is then no longer thermally-activated, 
we include it for two reasons. First, the MFPT is still mathematically
well-defined and can even be calculated analytically in the limit $\mu
\rightarrow \infty$. Second, for the bond rupture problem described in Refs.
\cite{z:evan97d} and \cite{z:izra97}, this regime has some interest, where it has been
called ``ultrafast". The time-ramped equation of motion becomes
$\pa_t x= \mu t$ with the  solution $x(t)=x+\mu t^2/2$.
This leads  to the exact asymptotic result
\beq
T(x,\mu)\approx \sqrt{2(x_r-x)/\mu}
\label{eq:ex}
\ee
and hence to $\alpha =1/2$ as mentioned above \cite{exact}.

For the SCCF approximation, the equation of motion
 becomes $\pa_t x= f$, which leads to $T(x)=(x_r-x)/f$. Replacing
$f$ by $\mu T(x)$ yields the  relation
\beq
T_s(x,\mu)\approx \sqrt{(x_r-x)/\mu}
\ee and hence $R_s(x,\mu) \approx \sqrt2/2$
for large $\mu$. In the same limit,  the adiabatic approximation
can be calculated by performing the integrals in (\ref{eq:ta}).
One obtains $R_a(x,\mu) \approx \sqrt{\pi}/2$
which is closer to the exact result than the SCCF approximation.

In an intermediate $\mu-$regime, the behavior can be extracted most easily from
the SCCF approximation. From the relations
$\ln T^K_s = \ln  T^K(0) - \mu T^K_s (x_s-x_m)$ one obtains
$
T_s^K\sim \ln \mu/\mu
$ and thus $\alpha =1$
 with logarithmic corrections. Expressed in terms of a mean rupture force
$f^*\equiv T_s^K \mu$, this behavior corresponds to
$f^*\sim \ln \mu$ \cite{z:evan97d}. As Fig. 1 shows,
the  data  show  a
large crossover with an effective exponent $\alpha \simeq  0.8$.

{\sl Conclusions --}
In this paper, we have analysed diffusive escape over a barrier
in the presence of a time-dependent force. We have derived an exact pde
for the mean first passage time and solved it both by numerical 
integration and by use of two approximations. Comparison of the (numerically)
exact solution with the adiabatic approximation introduced previously
shows that this approximation holds remarkably well for a
large range of loading rates.
Formally, our exact equation for the MFPT can easily
be  generalized  to include space-dependent
diffusion coefficients. Likewise, both
motion in higher dimensions and  inclusion of inertia terms (for smaller
friction) are amenable to the same treatment.
The numerical solution of the corresponding
equivalent of (\ref{eq:t2}), however, will become
 quite time-consuming. It will then
be helpful to explore the two approximations whose virtues we discussed
for our model case.

{\sl Acknowledgments --}
U.S. thanks the Canadian Institute of Advanced Research and M. Bloom
for the invitation
to participate at a workshop where E. Evans gave a talk on this problem.
Stimulating discussions with E. Evans, H. Gaub, M. Rief and M. Wortis are
gratefully acknowledged.
\appendix
\section{Small loading rates}

For small $\mu$, perturbation theory of the SCCF approximation
works out as follows. We first expand the  solution of (\ref{eq:ts}) as
\beq
T_0(x,f) = T^{(0)}(x) + f T^{(1)}(x) + O(f^2) .
\label{eq:p1}
\ee
 The first order term  obeys
$
\l T^{(1)}(x) = - \pa_x T^{(0)} (x) .$
We write its solution with the absorbing boundary conditions
as $
T^{(1)}(x) = \a [- \pa_x T^{(0)}] .
$ The linear operator $\a$ can be expressed easily by
quadratures and is formally the  inverse to $\l$. The final step
consists in replacing $f$ by $\mu T^{(0)}(x)= \mu \a[-1]$.
The first order result of the SCCF approximation thus reads
\beq
T_s(x,\mu) = T^{(0)}(x) +\mu \a[-1] \ \a[- \pa_x T^{(0)}] + O(\mu^2).
\label{eq:tap}
\ee A similar perturbative expansion for the adiabatic approximation
shows
that the term linear in $\mu$ coincides with the SCCF approximation
whereas the terms $O(\mu^2)$ differ in both approximations from each other.

Time (or rather $\tau$)-dependent perturbation theory in $\mu$ for the
solution of the exact equation
(\ref{eq:t2})
can be set
up as follows. Inserting the
 ansatz
\beq
T(x,\tau;\mu) = T^{(0)}(x) + \mu [T_0^{(1)}(x) + \tau T_1^{(1)}(x)]
 + O (\mu ^2)
\label{eq:p2}
\ee
in (\ref{eq:t2}), yields in $O(\mu,\tau^1)$ the equation
$\l T_1^{(1)} = -\pa_x T^{(0)}(x)$ with the solution
$T_1^{(1)}=\a[ -\pa_x T^{(0)}]$. Inserting this solution into the
equation  $O(\mu,\tau^0)$ yields $\l T_0^{(1)} = -T_1^{(1)}(x)$
with the solution $T_0^{(1)}(x) = \a[-\a[ -\pa_x T^{(0)}]]$. Thus, the
full equation leads to the perturbative result
\beq
T(x,\mu) = T^{(0)}(x) - \mu \a[\a[- \pa_x T^{(0)}]].
\label{eq:tp}
\ee
Since (\ref{eq:tp}) differs from (\ref{eq:tap}), neither approximation,
 perhaps somewhat surprisingly, reproduces the correct amplitude for the
term linear in $\mu$. Note, however, that the $\mu = 0$ value is
reproduced by both approximations. We suspect that by considering the
limit $Q \rightarrow \infty$ even the linear term shown in (\ref{eq:tap})
approaches the exact one shown in (\ref{eq:tp}), 
but we have not yet been able to show
this mathematically.

\eject
\begin {figure}[t]
\epsfig{figure=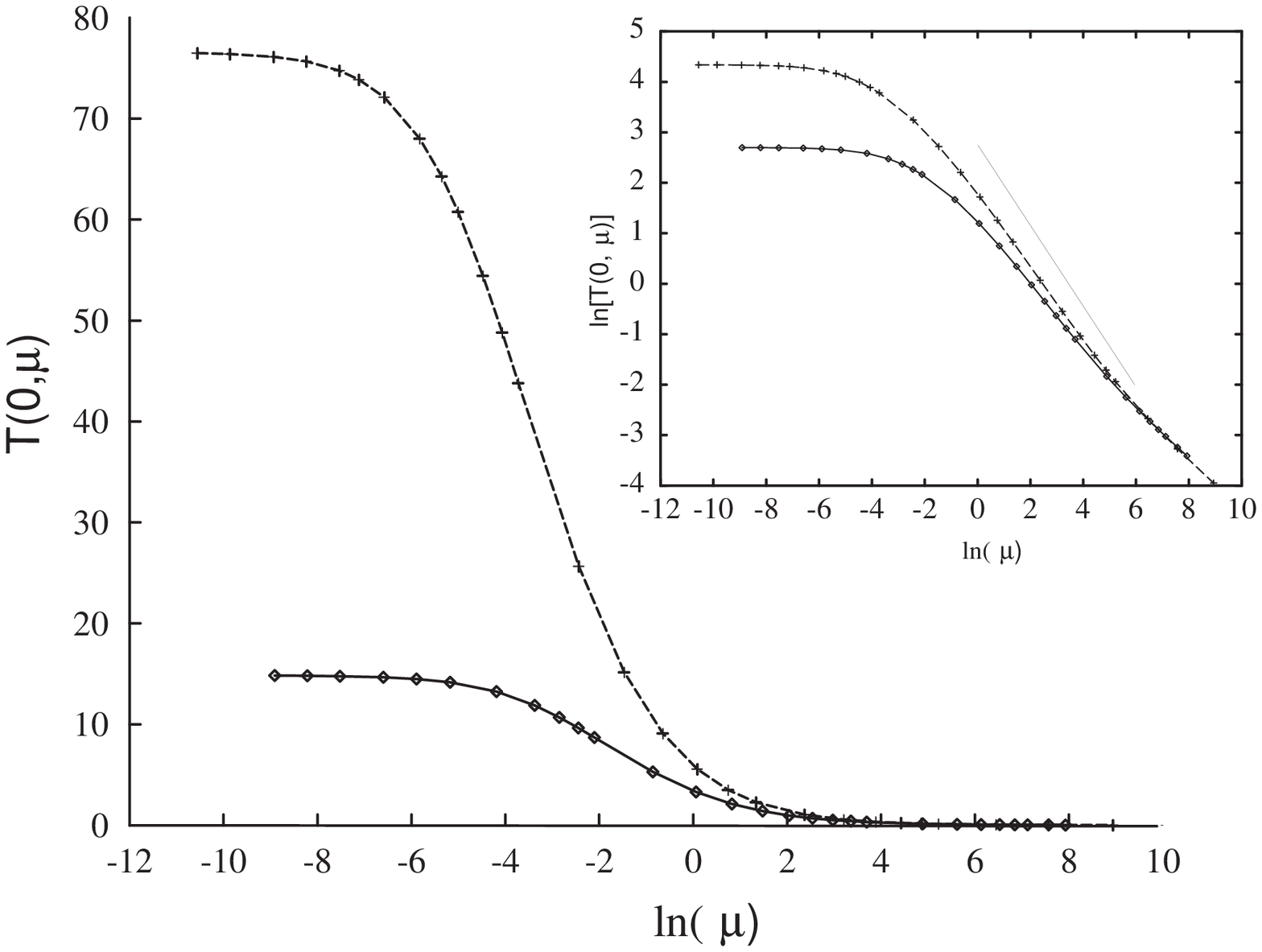,width=12cm }
\caption{Mean first passage  time $T(0,\mu)$
 as a function of loading rate $\mu$ for
$Q=7$ and $Q=5$ in the potential (\protect\ref{eq:V}). Inset shows the
same data logarithmically. For all $\mu$,
 the starting point has been chosen as $x=0$.
The absorbing boundary is $x_r =1.6$ and the (quite irrelevant) left
boundary is at $x_l=-1.6$.}
\label{fch1-2}
\end {figure} 
\begin {figure}[t]
\epsfig{figure=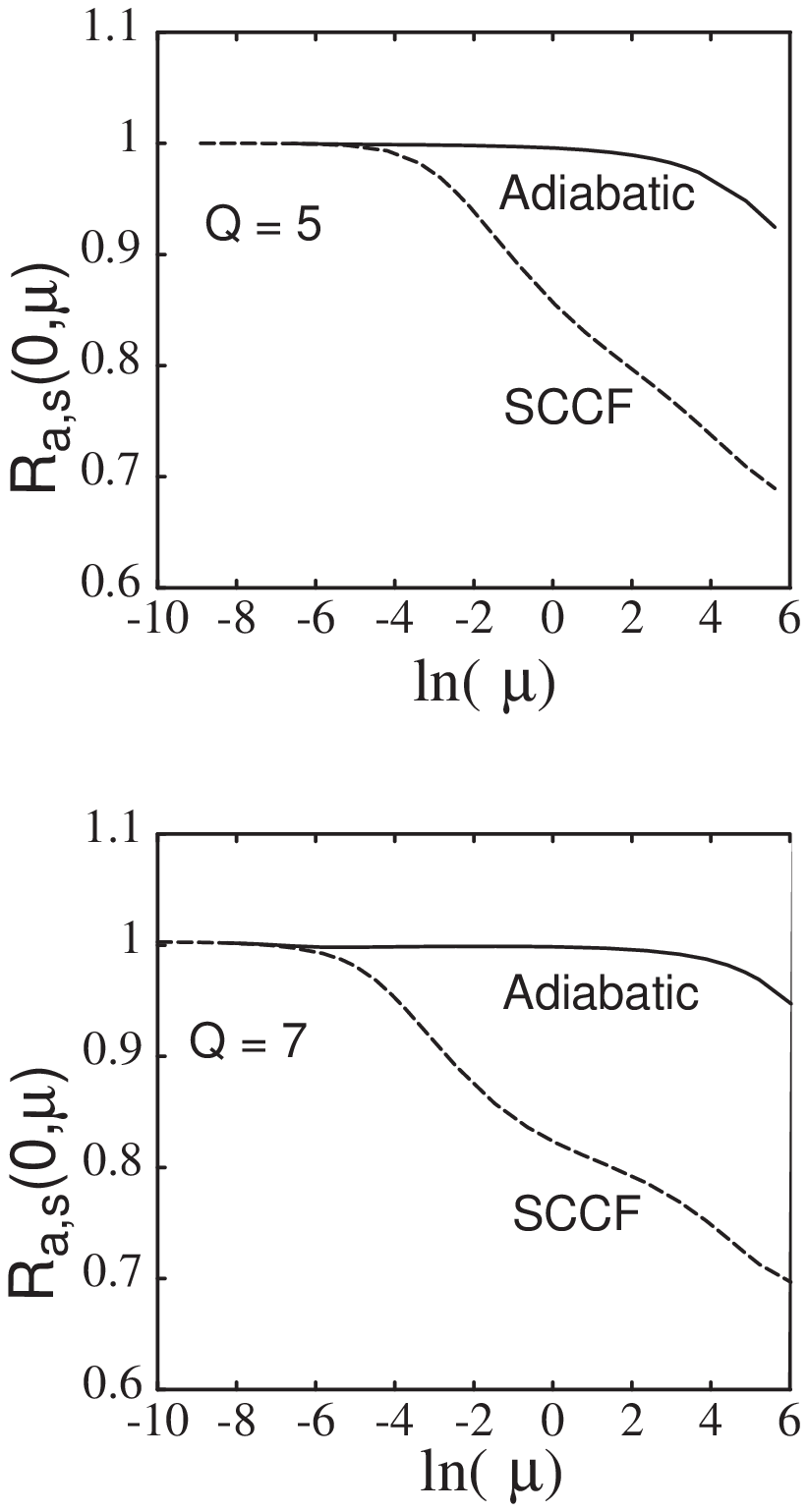,width=12cm }
\caption{Mean first passage  time $T(0,\mu)$
 as a function of loading rate $\mu$ for
$Q=7$ and $Q=5$ in the potential (\protect\ref{eq:V}). Inset shows the
same data logarithmically. For all $\mu$,
 the starting point has been chosen as $x=0$.
The absorbing boundary is $x_r =1.6$ and the (quite irrelevant) left
boundary is at $x_l=-1.6$.}
\label{fch1-2}
\end {figure} 


\begin{thebibliography}{10}
\bibitem{bc}
For a reflecting boundary at $x=x_l$, some trivial modifications of
subsequent formula will occur.
\bibitem{exact}
In this limit, the solution of Eq. \protect\ref{eq:t2} becomes
$T(x,\tau)= \sqrt{\tau^2 + 2(x_r-x)/\mu} -\tau $, since formally 
${\cal L}^\dagger=0$. Setting 
$\tau=0$ then 
also leads to 
Eq. \protect\ref{eq:ex}.
\end{thebibliography}
\end{document}